\documentclass[11pt,letterpaper]{amsart}
\usepackage{amsfonts}
\usepackage[english]{babel}
\usepackage{amssymb,amsthm,amsmath}
\usepackage[numbers,sort&compress]{natbib}
\usepackage{color}
\usepackage{eucal}
\usepackage{comment}
\usepackage{graphicx}
\usepackage[colorlinks,
            linkcolor=blue,
            anchorcolor=blue,
            citecolor=blue
            ]{hyperref}

\usepackage{fancyhdr}
\newcommand\shortitle{Entanglement entropy of free fermions in ergodic field: rigorous results and localisation-based methods.}
\numberwithin{equation}{section}

\def\Z{\mathbb{Z}}

\theoremstyle{plain}
\newtheorem{Theorem}{Theorem}
\theoremstyle{plain}
\newtheorem{Proposition}{Proposition}[section]

\newtheorem{Definition}[Proposition]{Definition}
\newtheorem{Lemma}[Proposition]{Lemma}
\newtheorem{aremark}[Proposition]{Remark}
\newenvironment{remark}{\begin{aremark}\rm}{\end{aremark}}

\newtheorem*{claim*}{Claim}
\newtheorem{Criterion}{Criterion}
\newtheorem{Remark}[Theorem]{Remark}
\newtheoremstyle{named}{}{}{\itshape}
{}{\bfseries}{.}{.5em}{\thmnote{#3}#1}
\theoremstyle{named}

\begin{document}

\title{Area laws for free fermions in ergodic field: spectral criteria and localisation-based methods}
\author{Leonid Pastur}
\address{Department of mathematics, King's College London, Strand, London
WC2R 2LS}
\email{leonid.pastur@kcl.ac.uk}
\address{Theoretical Department, B.Verkin Institute for Low Temperature
Physics and Engineering, 47 Nauky Avenue, Kharkiv, 61103, Ukraine}
\email{pastur@ilt.kharkov.ua }
\author{Mira Shamis}
\address{School of Mathematical Sciences, Queen Mary University of London,
Mile End Road, London E1 4NS, England}
\email{m.shamis@qmul.ac.uk}
\address{School of Mathematical Sciences, Holon Institute of Technology, 52 Golomb St.,
P.O.B. 305, Holon 5810201, Israel}
\email{miras@hit.ac.il}
\date{\today}

\begin{abstract}
We review rigorous results concerning the Area Law for the entanglement entropy of free fermions 
whose one-body Hamiltonian is an ergodic Schr\"odinger operator with a strongly localised pure point spectrum. 
Particular attention is paid to operators with deterministic potentials (quasi-periodic ones and those generated by 
subshifts of finite type). Our approach relies on reducing the problem to obtaining exponential 
estimates for the eigenfunction correlator. This provides an effective route from localisation 
to the Area Law and allows for the identification of additional spectral estimates that need to be proven 
for various classes of ergodic operators. We realise this approach by using various 
localisation techniques, including dynamical localisation and the uniformly localised eigenfunction 
property for bounded and unbounded ergodic operators (e.g. the Maryland model), 
as well as  a new method for obtaining the corresponding exponential bounds for 
Schr\"odinger operators generated by subshifts of finite type.
\end{abstract}

\maketitle

\qquad \textit{Dedicated to Uzy Smilansky on the occasion of his 85th birthday.}


\section{Introduction}

Quantum entanglement is one of the characteristic features distinguishing 
quantum systems from their classical counterparts. Its conceptual origins go back 
to the celebrated 1935 paper of Einstein, Podolsky, and Rosen, dealing with the 
completeness of quantum mechanics. Shortly afterwards, Schr\"odinger recognised 
the phenomenon discovered by Einstein et al. as a fundamental form of quantum 
correlation and introduced the term \emph{entanglement}. Since then, the subject has 
developed into an extensive area of research at the interface of quantum statistical mechanics, 
condensed-matter physics, quantum field theory, quantum gravity, 
quantum information theory, and related mathematical disciplines; 
see, for example, \cite{Al-Co:21,Ca-Co:11,Du:22,
Ei-Co:11,It-Ko:10,
La:15,Pe-Ei:09,Sc:19,Wi:18} and the references therein.

One of the widely used quantitative characteristics of entanglement between 
two parts of a quantum system is the \emph{entanglement entropy}. In the 
spatial setting considered in this survey, one separates a finite block $\Lambda$ 
of a quantum system occupying $\Z^d$ and studies the entanglement between $\Lambda$ 
and $\Z^d\setminus\Lambda$. The principal asymptotic problem is to 
determine the behaviour of the entanglement entropy $S_\Lambda$ as 
the size of $\Lambda$ tends to infinity. Thus, $\Lambda$ is viewed as a finite
but large (mesoscopic) part of an infinite quantum system, say  $\Lambda_{M} = [-M, M]^d,\, L = 2M + 1$. 

The principal asymptotic form of the entanglement entropy considered in this 
survey is the \emph{Area Law}. It is as follows
\begin{equation}
S_{\Lambda} =C^{\prime}L^{d-1}(1+o(1)),
\qquad L\rightarrow\infty ,
\label{eq:al}
\end{equation}
so that the leading term is proportional to the surface area of $\Lambda$. 
The Area Law is commonly associated with non-critical ground state of 
quantum system  and, 
in particular, with systems possessing a spectral gap above the ground state, 
although the precise hypotheses under which it holds depend strongly on the 
class of systems under consideration.

For comparison, we also recall the \emph{Enhanced Area Law}
\begin{equation}
S_{\Lambda} =C^{\prime\prime}L^{d-1}\log L\,(1+o(1)),
\qquad L\rightarrow\infty .
\label{eq:eal}
\end{equation}
The logarithmic enhancement occurs in important gapless or critical 
systems \cite{Ca-Co:11}. Although \eqref{eq:al} and \eqref{eq:eal} differ only 
by a logarithmic factor, their rigorous derivations may require substantially 
different information about the system in question. The Enhanced Area 
Law will be discussed only briefly, mainly in order to contrast it with 
the localised regime that is the principal subject of this survey.

For completeness, we also mention the \emph{Volume Law}
\begin{equation*}
S_{\Lambda} =C^{\prime\prime\prime}L^{d}(1+o(1)),
\qquad L\rightarrow\infty ,
\end{equation*}
which occurs in substantially different settings, such as mixed states 
at non-zero temperature or sufficiently highly excited pure states, and will not be considered below.

Note that the large size asymptotic form
\[
F_\Lambda=f_{\text{vol}}L^{d}+f_{\text{area}}L^{d-1}+o(L^{d-1}), \; L \to \infty
\] 
of an extensive quantity $F_\Lambda$ is well known in statistical mechanics and condensed 
matter physics. It is also the case in the mathematical but related context of Szeg\H{o}'s
theorem with a regular symbol \cite{Si:11}. A novel feature arising from quantum 
mechanics is the absence of the volume (bulk) contribution $f_{\text{vol}}$  in the 
Area Law \eqref{eq:al} and the Enhanced Area Law \eqref{eq:eal}.

The proofs of the Area Law and the Enhanced Area Law are highly non-trivial 
for general multidimensional and interacting quantum systems, see e.g. 
\cite{Ca-Co:11,Ei-Co:11,Go-Ha:10,It-Ko:10,La:15,Pe-Ei:09,Wi:18}. This has led to considerable 
interest in free fermions, which retain non-trivial entanglement properties while allowing an exact 
reduction of the problem to spectral analysis of the one-body Hamiltonian of free fermions. The reduction makes the problem accessible, but the resulting asymptotic analysis 
remains delicate: the entropy is a highly non-linear and non-smooth functional of the one-body Hamiltonian.


The translation invariant case has been rigorously studied in great detail, where the Enhanced Area Law \eqref{eq:eal} holds (see \cite{La-Wi:80,Le-Co:14,Le-Co:17,So:17} and the references therein). {Consequently, the problem of the asymptotic behavior of entanglement entropy in inhomogeneous extended (disordered, incommensurate, etc.) systems arose naturally, much as it did in statistical physics and condensed matter physics. 

The problem was considered in \cite{EPS,Pa-Sl:18} under the 
assumption that
$H$ is the random  Schr\"odinger operator.} 
It was shown that sufficiently strong exponential decay of the spectral (Fermi) projection 
of $H$ inherent in its strong localisation regime yields the Area Law for the expectation 
of the entanglement entropy. However, a careful examination of the work 
shows that its results are not intrinsically restricted to random operators: 
they apply to any ergodic finite-difference operator for which the required 
exponential decay estimate for the Fermi projection is available.

The subsequent development of this approach, especially a presentation 
of results obtained via its reformulation in terms of the so-called eigenfunction correlator, 
is one of the central subjects of the present survey. The Fermi projection remains the spectral 
object entering the definition of the entropy, whereas the eigenfunction correlator is 
often a more convenient quantity to deal with. Its decay simultaneously controls the Fermi projection 
and exponential dynamical localisation (see, e.g. \cite{AW}) and therefore provides a workable route 
from strong spectral localisation to the Area Law. This formulation describes the mechanism behind 
the presently available rigorous results; it does not assert that localisation is 
necessary for an Area Law. It is also valid, for instance, in any internal gap of the spectrum of the one-body Hamiltonian.

The paper is organised as follows. In Section~\ref{s:prob}, 
we formulate the corresponding asymptotic problem  with  and describe 
its reduction to the problem of the spectral analysis of ergodic operators. 
In Section~\ref{sec:tools} we outline the spectral tools both known and new to be used in the analysis. 
Section~\ref{sec:results} contains the description of recent results
on the asymptotic form (Area Law) for the entanglement entropy of free fermions 
having certain finite difference operators with dynamically generated ergodic coefficients. 

We will mostly focus on two basic operators but also discuss a 
collection of the related operators. Our basic operators are the multidimensional 
Maryland model and the one-dimensional discrete Schr\"odinger operators whose 
potential is generated by the subshift of finite type.    

The Maryland model, since its introduction in the early 1980s, has remained an 
important testing ground of localisation theory for ergodic, specifically  quasiperiodic, 
operators, see \cite{Pa-Fi:92}, Chapter 18 and \cite{Si:85} for initial basic 
results and \cite{Ji-Li:17,JY,HJY,FK,FS} for certain subsequent developments.  
In particular, the spatial behaviour of eigenfunctions has been studied extensively.  
The exponential localisation estimates for the eigenfunctions obtained in \cite{JY,HJY} 
can be made uniform in spectral parameter 
varying in a 
fixed compact interval, hence, cannot be used for the whole unbounded 
spectrum of the Maryland model.  
In \cite{PS}, the uniform exponential bounds for eigenfunctions (ULE) were established 
for the whole 
 unbounded pure-point spectrum: 
on the whole real line in the multidimensional model, and on the 
two semi-infinite pure-point components in the one-dimensional case, which together cover the whole
line when the singular-continuous interval is absent.

Another basic class of operator consists of the one-dimensional discrete  Schr\"odinger 
operator with potential generated by subshifts of finite type. 
It takes us far beyond quasiperiodic operators generated by non-mixing irrational rotations, 
namely, towards random operators with Markov potential, while remaining entirely deterministic. 
This is because the corresponding dynamics is exponentially mixing, so that deterministic 
constraints coexist with rapid decay of correlations. They thus provide a natural 
intermediate regime between quasiperiodic and random operators. The exponential 
decay of the eigenfunction correlator, and consequently the Area Law, 
was established for this class in \cite{PS}.  The spectral results of \cite{ADZ} 
form a quite important input, but obtaining eigenfunction-correlator 
decay requires an essentially new additional mechanism developed in \cite{PS}.

{
We conclude this section with a remark concerning our bibliography. The literature on quantum entanglement is extremely extensive even when restricted to works dealing with free fermions. Consequently, we have limited our references to monographs, review articles, and rigorous results or technical papers that are directly related to the Area Law for free fermions. It should be noted that even this collection of references is far from exhaustive.}

\section{Problem}\label{s:prob}
We now turn to the mathematical formulation of the problem. Throughout 
this survey we consider free lattice fermions and define the corresponding 
entanglement entropy explicitly in terms of the Fermi projection of the associated one-body Hamiltonian.

A system of free lattice fermions is completely characterised by its one-body Hamiltonian $H$, a self-adjoint operator acting in $\ell^2(\mathbb Z^d)$
\begin{equation}
H:=\{H(\mathbf m,\mathbf n)\}_{\mathbf m,\mathbf n\in\mathbb Z^d},
\qquad
\overline{H(\mathbf m,\mathbf n)} = H(\mathbf n,\mathbf m).
\label{eq:H}
\end{equation}

One of the principal examples is the discrete Schr\"odinger operator
\begin{equation}
\begin{split}
(H\psi)(\mathbf n)
&=
(\Delta\psi)(\mathbf n)+(V\psi)(\mathbf n)
\\
&=
\sum_{|\mathbf n-\mathbf m|=1}\psi(\mathbf m)
+
V(\mathbf n)\psi(\mathbf n),
\qquad
\mathbf n\in\mathbb Z^d,
\end{split}
\label{eq:Sch}
\end{equation}
where $\Delta$ is the discrete Laplacian and $V$ is a real-valued multiplication operator. 

More generally, we will confine ourselves to
the finite-difference operators
\begin{equation}
H_\omega = W+V= \{H(\mathbf m,\mathbf n)\}_{\mathbf m,\mathbf n\in\mathbb Z^d},
\label{eq:lro-a}
\end{equation}
with
\begin{equation}
H(\mathbf m,\mathbf n) = W(\mathbf m-\mathbf n)
+
V(\mathbf n)\delta_{\mathbf m,\mathbf n}.
\label{eq:lro}
\end{equation}
where the convolution (hopping) term $W$ satisfies
\begin{align}
& \hspace{2cm} W(-\mathbf n)=\overline{W(\mathbf n)}, \notag
\\
&|W(\mathbf n)|
\leq
C_W e^{-\rho|\mathbf n|},
\quad
C_W<\infty,\quad \rho>0,
\label{eq:W}
\end{align}
and
\[
|\mathbf n|=|n_1|+\cdots+|n_d|.
\]
The Schr\"odinger operator \eqref{eq:Sch} is the particular case in which $W=\Delta$.

Let $\sigma(H)$ be the spectrum of $H$ and 
\begin{equation}\label{eq:edges}
\varepsilon_-=\min_{\lambda \in \sigma(H)}\lambda, \qquad 
\varepsilon_+=\max_{\lambda \in \sigma(H)}\lambda
\end{equation}
 be its (possibly infinite) lower and upper endpoints.

Let $\mathcal E_H(d\lambda)$ be the resolution of the identity of $H$. 
The Fermi projection at the Fermi energy $\varepsilon_F \ge \varepsilon_-$ is
\begin{equation}
P(\varepsilon_F):=\mathcal E_H([\varepsilon_-,\varepsilon_F])
 = \chi_{[\varepsilon_-,\varepsilon_F]}(H),
\label{eq:FP}
\end{equation}
where $\chi _{I}$ is the indicator of $I \subset \mathbb{R}$.

For a finite block $\Lambda\subset\mathbb Z^d$, let
\begin{equation}
P_\Lambda(\varepsilon_F) = \chi_\Lambda
P(\varepsilon_F)
\chi_\Lambda
=
P(\varepsilon_F)|_\Lambda
\label{eq:pla}
\end{equation}
be the restriction of the Fermi projection to $\ell^2(\Lambda)$, where
\begin{equation}
\chi_\Lambda:
\ell^2(\mathbb Z^d)
\rightarrow
\ell^2(\Lambda)
\label{eq:chi}
\end{equation}
is the coordinate projection. 

The entanglement entropy of the corresponding free-fermion ground state is
\begin{equation}
S_\Lambda(\varepsilon_F) =\operatorname{Tr}_\Lambda
h\bigl(P_\Lambda(\varepsilon_F)\bigr),
\label{eq:ee}
\end{equation}
where $\operatorname{Tr}_\Lambda$ denotes the trace in $\ell^2(\Lambda)$ and
\begin{equation}
h(x)= -x\log x-(1-x)\log(1-x),
\qquad
x\in[0,1],
\label{eq:h}
\end{equation}
is the binary entropy function, see \cite{Ei-Co:11,Pe-Ei:09}.

The above formulas 
reduce the initial many-body problem  to a problem of spectral theory. 
The entropy is determined entirely by the restriction of the Fermi projection of the one-body Hamiltonian 
to the finite block, but its dependence on that restriction is non-linear and non-smooth. Consequently, 
even when the Fermi projection is known explicitly, extracting the large-$\Lambda$ asymptotics 
of \eqref{eq:ee} may remain a difficult problem.

If $\varepsilon_F \le \varepsilon_-$, then $P(\varepsilon_F)=0$, and if $\varepsilon_F>\varepsilon_+$, then $P(\varepsilon_F)=I$. In either case,
$
S_\Lambda(\varepsilon_F)=0.
$
We shall therefore assume  that
\begin{equation}
\varepsilon_F\in (\varepsilon_-,\varepsilon_+),
\label{eq:kap}
\end{equation}
so that $\varepsilon_F$ belongs either to the spectrum of $H$ or to an internal spectral gap.

To study the asymptotic form of $S_\Lambda$ as $\Lambda$ grows, one needs a class of 
Hamiltonians whose large-scale spatial structure is ``sufficiently homogeneous'' to provide a 
regular asymptotic behaviour. The natural framework is given by ergodic operators
\cite{Pa-Fi:92}. 

Let 
\begin{equation}\label{eq:psp}
(\Omega,\mathcal F,\mathbf P)
\end{equation}
be a probability space equipped with a $\mathbf{P}$-preserving ergodic action
\begin{equation}\label{eq:auto}
\{T^{\mathbf n}:\Omega\to\Omega\}_{\mathbf n\in\mathbb Z^d}.
\end{equation}
A map from $\Omega$ to the set of selfadjoint
operators in $\ell^2(\mathbb{Z}^d)$
\begin{equation}\label{eq:H-omega}
\{H_\omega\}_{\omega \in \Omega}, \;\; H_\omega=\{H_{\omega}(\mathbf m,\mathbf n)\}_{\mathbf m,\mathbf n\in\mathbb Z^d}
\end{equation}
is called an \emph{ergodic operator} if it is covariant under lattice translations:
\begin{equation}
H_{T^{\mathbf k}\omega}(\mathbf m,\mathbf n)=
H_{\omega}(\mathbf m+\mathbf k,\mathbf n+\mathbf k),
\qquad
\mathbf k,\mathbf m,\mathbf n\in\mathbb Z^d.
\label{eq:herg}
\end{equation}
Equivalently, if $U_{\mathbf k}$ denotes the lattice shift,
\begin{equation}\label{eq:Uk}
(U_{\mathbf k}\psi)(\mathbf n)=
\psi(\mathbf n+\mathbf k),\quad \psi\in\ell^2(\Z^d),
\end{equation}
then
\begin{equation}\label{eq:HUHU}
H_{T^{\mathbf k}\omega}=U_{\mathbf k}H_\omega U_{\mathbf k}^{-1}.
\end{equation}
Thus, translating the operator in physical space corresponds to moving 
the point $\omega \in \Omega$ along the orbit of the underlying dynamical system
\eqref{eq:psp} -- \eqref{eq:auto}.

For the finite-difference operators \eqref{eq:lro-a} -- \eqref{eq:W}, including 
the Schr\"odinger operator \eqref{eq:Sch}, the ergodicity property \eqref{eq:herg} means 
that the potential is generated by sampling along  an orbit:
\begin{equation}
V_\omega(\mathbf n)= V(\omega,\mathbf n) = v(T^{\mathbf n}\omega),
\qquad
\mathbf n\in\mathbb Z^d,
\quad
\omega\in\Omega,
\label{eq:epotg}
\end{equation}
where $v:\Omega\to\mathbb R$ is a sampling function. This single 
formulation includes random, quasiperiodic, limit-periodic, and other dynamically generated potentials, 
including those generated by the so-called subshifts of finite type, see e.g. \cite{Br-St:02}. Thus, 
once an ergodic dynamical system and a sampling function are specified, formula \eqref{eq:epotg} 
defines the corresponding ergodic potential. Ergodicity, rather than just
randomness, 
is therefore the structural assumption relevant to the large-volume asymptotic regime considered in this survey.

The simplest class of ergodic operators are convolution (translation invariant) operators. 
In this case the probability space consists of a single point and
\begin{equation}
H(\mathbf m,\mathbf n)=W(\mathbf m-\mathbf n),
\qquad
\sum_{\mathbf n\in\mathbb Z^d}|W(\mathbf n)|<\infty .
\label{eq:con}
\end{equation}
These operators are diagonalised by the Fourier transformation. In particular,
for $d=1$  the matrix elements of the corresponding Fermi projection 
\eqref{eq:FP} have the simple explicit form
\begin{equation}
P(m,n)=\frac{\sin k_F(m-n)}{\pi(m-n)},\qquad m, n\in\Z,
\label{eq:FP1-con}
\end{equation}
where $k_F > 0$ is determined by the symbol (Fourier transform) $w$ of $W$
via $w(k_F) = \varepsilon_F$. 

The slow, oscillatory decay in \eqref{eq:FP1-con} is characteristic of an extended regime 
(absolutely continuous spectrum) of $H$. For broad classes of translation invariant and 
periodic operators, a highly non-trivial asymptotic analysis leads to the Enhanced Area Law \eqref{eq:eal}, see
\cite{La-Wi:80,Le-Co:14,Le-Co:17,Pf-So:18,So:17} and the references therein.

This translation invariant setting will serve mainly as a point of comparison. 
Our principal concern is the Area Law \eqref{eq:al} for free fermions whose 
one-body Hamiltonian is an ergodic finite-difference operator with strong spectral  (Anderson) localisation,
an important quantitative form of the pure point spectrum.

The apparent contrast between translation invariant and ``strongly'' localised ergodic  cases 
(Enhanced Area Law versus Area Law) suggests certain links   between the asymptotic behaviour 
of the entanglement entropy and the spectral properties of the one-body Hamiltonian. 
In the translation invariant case, the spectrum is typically absolutely continuous and the 
generalised eigenfunctions are extended (plane waves). The corresponding Fermi 
projection has slowly decaying matrix elements \eqref{eq:FP1-con}, and the 
resulting entropy typically obeys the Enhanced Area Law. In the ``strongly'' localised regime by contrast
the spectrum is typically pure point and the eigenfunctions are exponentially localised.  The corresponding
Fermi projection has exponentially decaying matrix elements \eqref{eq:exp-decay} and 
the entanglement entropy obeys the Area Law \eqref{eq:al}.

This comparison is useful, but it should not be interpreted as a general correspondence 
between spectral type and the asymptotic form of the entanglement
entropy  $S_\Lambda(\varepsilon_F)$. Spectral type alone does not determine 
the asymptotic form of  $S_\Lambda(\varepsilon_F)$ at a fixed Fermi energy.
This is because the form depends on the spatial decay of the matrix elements of 
the Fermi projection in an infinitesimal  neighbourhood of the Fermi energy, 
whereas the decomposition of the spectrum into absolutely continuous, singular 
continuous, and pure point components describes the operator on spectral 
sets of fixed size. The distinction is not merely formal: random operators with
 pure point spectrum may exhibit the Enhanced Area Law at isolated Fermi 
energies \cite{Mu-Co:20}. Thus, the relevant input is not the spectral type by itself, 
but quantitative information on the spatial behaviour of the Fermi
projections.

This observation also explains why pure point spectrum, or even exponential decay of individual eigenfunctions, 
does not by itself provide the quantitative input used in the presently available proofs of the Area Law. Such spectral 
information may be sufficient in a broader theory that has not yet been developed; the existing arguments require 
uniform or averaged estimates controlling the Fermi projection, or a spectral quantity that dominates it. 
Whether substantially weaker spectral assumptions may still imply an Area Law remains open.

\section{Spectral Tools}\label{sec:tools}

It was explained above that the presently known rigorous proofs of
the Area Law for free fermions rely on the reduction of the  problem to suitable quantitative localisation estimates 
for the underlying one-body Hamiltonian. In this section we 
formulate a general criterion making this reduction explicit. The criterion is a reformulation in 
terms of the eigenfunction correlator of the criterion given in \cite{EPS} in terms of the 
Fermi projection. We then formulate  a finite-volume criterion for the exponential decay
of the eigenfunction correlator and finally recall the notion of the Lyapunov
exponent, which will appear in the subsequent discussion of quasiperiodic
and symbolic models.

The mathematical strategy underlying the known results on the entanglement
entropy separates naturally into two steps. 

The first is a problem in spectral theory: one must obtain sufficiently precise information on the entries of
the Fermi projection \eqref{eq:FP} of
$H_\omega$\begin{equation}
P_\omega(\varepsilon_F) = \chi_{(\varepsilon_-,\varepsilon_F]}(H_\omega) 
= \{P_\omega(\mathbf m,\mathbf n;\varepsilon_F)\}_{\mathbf m,\mathbf n\in\mathbb Z^d},
\label{eq:FP-ent}
\end{equation}
or on a more tractable spectral quantity that dominates it. 

The second step is the asymptotic analysis of the non-linear and non-smooth
functional
\[
\operatorname{Tr}_\Lambda
h\bigl(P_\omega(\varepsilon_F)|_{\Lambda}\bigr)
\]
of $H$ as the block $\Lambda$ tends to infinity.

The relative difficulty of these steps depends strongly on the operator under consideration. 
For translation-invariant operators, the Fermi projection can be easily obtained  (see e.g. \eqref{eq:FP1-con}), 
while the derivation of entropy asymptotics requires a highly non-trivial analysis of truncated convolution 
and certain Wiener--Hopf operators; see \cite{La-Wi:80,Le-Co:14,Le-Co:17,So:17}. For strongly localised ergodic 
operators, by contrast, obtaining that estimate (see e.g. \eqref{eq:exp-decay}) is the principal difficulty. 
Once an appropriate exponential decay estimate for the Fermi projection has been established, 
the  asymptotic  formula for the entanglement entropy follows from the result  of  \cite{EPS}.
Indeed, although the proof in \cite{EPS} is presented for random Schr\"odinger operators, 
its entropy part uses only ergodicity and the decay estimate \eqref{eq:exp-decay}. 

The relevant decay estimate on \eqref{eq:FP-ent} is
\begin{equation}
\mathbf E
\bigl\{
|P_\omega(\mathbf m,\mathbf n;\varepsilon_F)|
\bigr\}
\leq
C e^{-c|\mathbf m-\mathbf n|},
\qquad
\mathbf m,\mathbf n\in\mathbb Z^d,
\label{eq:exp-decay}
\end{equation}
where $\mathbf E\{\cdot\}$ denotes the expectation with respect to $\mathbf P$ of \eqref{eq:psp} and 
$C<\infty$, $c>0$ are independent of $\mathbf m$ and $\mathbf n$, although they may depend on the Fermi energy. 

Such estimates are always available in spectral gaps by the general Combes--Thomas bound and contour integration, see e.g. \cite{AW}. 
They are also available in ``strongly'' localised parts of the spectrum for several classes of random ergodic operators and are considerably stronger than the statement that the spectrum is pure point, 
see \cite{AW,Fr-Sp:83,St:11}.

More precisely, it is proved in \cite{EPS} that if
\[
\Lambda_M=[-M,M]^d\cap\mathbb Z^d,
\qquad
L=2M+1,
\]
and \eqref{eq:exp-decay} holds at the Fermi energy $\varepsilon_F$, then
we have\begin{equation}
0<
\lim_{L\to\infty}
L^{-(d-1)}
\mathbf E
\bigl\{
S_{\Lambda_M}(\varepsilon_F)
\bigr\} := s_d(\varepsilon_F)
<
\infty,
\label{eq:area-law-criterion}
\end{equation}
with a certain explicit form of  $s_d(\varepsilon_F)$.

In other words, the implication
\emph{Bound \eqref{eq:exp-decay} $\Longrightarrow$ Area Law \eqref{eq:area-law-criterion}}
applies to any ergodic finite-difference operator satisfying 
\eqref{eq:exp-decay}.
Hence,
the principal model-dependent problem is the proof of \eqref{eq:exp-decay}.

However, the direct estimates of the Fermi projection, being quite efficient for random operators \cite{AW,St:11},
often are not available for
non-random ergodic operators. A more appropriate quantity 
here proved to be the \emph{eigenfunction correlator}, introduced by Aizenman \cite{A94} 
and widely used in localisation theory \cite{AW,St:11}. 

Let $H$ be a self-adjoint operator in $\ell^2(\mathbb Z^d)$ and let $I\subset\mathbb R$ be a spectral set. 
Then the eigenfunction correlator is defined as
\begin{equation}
Q_I(\mathbf m,\mathbf n)
:=
\sup_{\substack{\phi\in C(\mathbb R),\, \|\phi\|_\infty\leq 1}}
\left|
\left\langle
\delta_{\mathbf m},
\phi(H)\chi_I(H)\delta_{\mathbf n}
\right\rangle
\right|, \; \; \mathbf m,\mathbf n\in\mathbb Z^d.
\label{eq:ef-cor}
\end{equation}
If the spectrum of $H$ in $I$ is pure point, with eigenvalues
$\{\lambda_l\}_{l}$ and corresponding normalised eigenfunctions
$\{\psi_l\}_{l}$, then
we have\begin{equation}
Q_I(\mathbf m,\mathbf n) = \sum_{\lambda_l\in I}
|\psi_l(\mathbf m)|
\,|\psi_l(\mathbf n)|.
\label{eq:ef-cor1}
\end{equation}
This representation explains both the terminology and the usefulness of the eigenfunction correlator: 
it records the collective spatial behaviour of all eigenfunctions with eigenvalues in $I$, 
rather than the behaviour of a single eigenfunction. {This difference is especially non-trivial for ergodic operators whose spectrum
is dense. }

The eigenfunction correlator controls the Fermi projection $P_\omega$ directly,
since we have by the spectral theorem for $P_\omega$ and \eqref{eq:ef-cor}\begin{equation}
|P_\omega(\mathbf m,\mathbf n;\varepsilon_F)|
\leq
Q_{(\varepsilon_-,\varepsilon_F]}
(\mathbf m,\mathbf n).
\label{eq:PQ}
\end{equation}
 It is noteworthy that $Q_I$\  also controls an important characteristic
of the localisation and the time evolution for a spectral set $I$:
\begin{equation}
\sup_{t\in\mathbb R}
\left|
\left\langle
\delta_{\mathbf m},
e^{-itH}\chi_I(H)\delta_{\mathbf n}
\right\rangle
\right|
\leq
Q_I(\mathbf m,\mathbf n).
\label{eq:evQ}
\end{equation}

The bound \eqref{eq:PQ} implies the following reformulation of the
criterion
of \cite{EPS},  based on \eqref{eq:exp-decay}.

\begin{Criterion}[Eigenfunction correlator]
\label{cr:EC}
Assume that the one-body Hamiltonian of free fermions is an ergodic operator
such that its eigenfunction correlator \eqref{eq:ef-cor} admits the
estimate
\begin{equation}
\mathbf E
\bigl\{
Q_I(\mathbf m,\mathbf n)
\bigr\}
\leq
C e^{-c|\mathbf m-\mathbf n|},
\quad
\mathbf m,\mathbf n\in\mathbb Z^d,  \; C < \infty, \; c>0.
\label{eq:ef-cor-decay}
\end{equation} 
on the interval
 \begin{equation}\label{eq:Ient}
I=(\varepsilon_-,\varepsilon_0],
\end{equation}
where $\varepsilon_-$ is the bottom of the spectrum \eqref{eq:edges},    
$I$ contains a spectral region,  $\varepsilon_F$ is in the region or is to the right of it.

Then the
expectation of the corresponding entanglement entropy obeys the Area
Law \eqref{eq:area-law-criterion}, cf. \eqref{eq:al}.
\end{Criterion}

By the way,  combining \eqref{eq:evQ} and \eqref{eq:ef-cor-decay}, we obtain  
the property of the exponential dynamical localisation in expectation on $I$ (cf. \eqref{eq:evQ}):
\begin{equation}  \label{eq:exp-dyn-loc}
\mathbf{E}\Big\{\sup_{t\in\mathbb{R}}|\langle \delta_\mathbf{m},
e^{-itH_\omega}\chi_I (H_\omega)\delta_\mathbf{n}|\Big\} \leq C e^{-c|%
\mathbf{m}-\mathbf{n} |}.
\end{equation}

The advantage of \eqref{eq:ef-cor-decay} is methodological rather than conceptual: 
the Fermi projection remains a basic quantity from which the entropy is defined, while the eigenfunction 
correlator is often more accessible to localisation techniques. It therefore provides a common 
language in which results obtained by very different spectral methods can be converted into Area Laws. 

The framework described above was developed  in the work \cite{PS} of authors, 
where it was applied to broad classes of deterministic  ergodic finite-difference operators.  
It was found that depending on the model, this may be achieved by different localisation mechanisms.
In certain cases, the required decay estimates were already implicit in existing localisation results 
and had to be just extracted and/or reformulated in an appropriate form. In other cases, the available spectral 
theory did not provide the necessary quantitative control and had to be substantially extended. 
In doing so, we will essentially employ the following ingredients of the spectral theory of ergodic operators:

\medskip

(i) the uniformly localised eigenfunction property (ULE),

\smallskip
(ii) exponential dynamical localisation,

\smallskip
(iii) finite-volume Green-function estimates.

\smallskip \noindent
We will briefly discuss the meaning of (i) and (iii) below and use them in 
Sections~\ref{subsec:maryland} and~\ref{subsec:subshift}. 
Item (ii) is recalled and used in Section~\ref{subsec:qp}.
 
\medskip \noindent
\textbf{Uniformly localised eigenfunctions}. The property  is as follows \cite{dRJLS}. 
Let $H$ be a self-adjoint operator in $\ell^2(\mathbb Z^d)$ which
has the simple pure point spectrum on $I\subseteq \mathbb{R}$ with 
eigenvalues $\lambda_l \in I$ and the corresponding eigenfunctions
\[
\{\psi_l\}_{\lambda_l \in I},\quad
\psi_l=\{\psi_l(\mathbf n)\}_{\mathbf n\in\mathbb Z^d}.
\] 
We say that $H$ has \emph{uniformly localised eigenfunctions} (ULE) on $I$, if there
exist  
$C<\infty$ $c>0$ which do not depend on $l, \; \lambda_l \in I$ and such that
\begin{equation}\label{eq:lc}
|\psi_l(\mathbf n)|
\le
Ce^{-c|\mathbf n-\mathbf n_l|},
\qquad
\mathbf n\in\mathbb Z^d,
\end{equation}
where the points $\{\mathbf n_l\in\mathbb Z^d\}_l$ are known as the localisation centres of the
corresponding  eigenfunctions and may depend on $H$.

This is a mathematical realisation of the quantum-mechanical intuition that, 
if the potential is a sufficiently complex (disordered) function, then there are no quantum 
numbers other than the coordinate $\mathbf{n}_l$ for indexing the wave functions.

ULE is a considerably stronger property than pure point spectrum or
even exponential localisation of individual eigenfunctions. The property proves to be fairly subtle. 
In particular, it does not hold for the discrete Schr\"odinger equation with
random potential  in any dimension or for the almost Mathieu operator \eqref{eq:AMO}, see \cite{dRJLS}. 

Moreover, even if a certain operator possesses this property, the proof of estimate \eqref{eq:ef-cor-decay}, 
and, consequently, the applicability of Criterion \ref{cr:EC}, is not immediate. Indeed,
the ULE estimate \eqref{eq:lc} provides the exponential decay of every term in the sum in the r.h.s. of 
formula \eqref{eq:ef-cor1}  for the eigenfunction correlator $Q_I$, and to obtain a bound for the sum, one 
has to control the distribution of localisation centres. In addition, the parameters 
$C$, $c$, and $\{\mathbf n_l\}_{l}$ in \eqref{eq:lc}  may
depend on the operator in question. Thus, for ergodic operators \eqref{eq:psp}
-- \eqref{eq:HUHU}
these parameters
may depend on  $\omega$ and this must be taken into account when applying the ULE property 
to prove an important  bound  \eqref{eq:ef-cor-decay} for such operators.

\medskip \noindent

\textbf{Finite-volume Green-function estimates}.
Let
\[
M=\{M(j,k)\}_{j,k=-n}^{n}
\]
be a $(2n+1)\times(2n+1)$ matrix and denote by
\[
G_M(j,k;\lambda)
=
(M-\lambda)^{-1}(j,k)
\]
its 
Green function. We shall be interested in
those spectral parameters for which the Green function connecting the centre
of the interval to one of its end points is not too small.

\begin{Definition}
For $\epsilon>0$, define
\begin{equation}\label{eq:bad}
\begin{split}
\mathrm{Bad}(M,\epsilon)
=
\Bigl\{
\lambda\in\mathbb R:\,
\max\bigl(
|G_M(0,n; \lambda)|,\,
|G_M(0,-n; \lambda)|
\bigr)
\ge\epsilon
\Bigr\},
\end{split}
\end{equation}
the corresponding set of ``bad'' spectral parameters.
\end{Definition}

The use of such "finite-volume" Green functions, where $M$ is the restriction
of the Hamiltonian $H$ to a finite domain of $\mathbb{Z}^d$,  as a tool for proving
localisation goes back to the pioneering work of
Fr\"ohlich and Spencer \cite{Fr-Sp:83} and has since become one of the
main ingredients of the modern theory of Anderson localisation known as the
multiscale analysis, see  \cite{AW,St:11}. The following criterion, proved in
\cite{PS}, provides a convenient bridge between the estimates of  finite-volume 
Green functions  of and exponential decay of the eigenfunction correlator.
Here $I\subset\mathbb R$ is an arbitrary interval.

\begin{Lemma}[Finite-volume criterion for the eigenfunction correlator]
\label{lem:bad-res}
Let $H$ be a bounded ergodic one-dimensional discrete
Schr\"odinger operator \eqref{eq:Sch}, \eqref{eq:herg}--\eqref{eq:epotg}. Assume that for any $n
\geq n_0$ and  $|k|, |l| \leq n^2$ such that $|k - l| \geq 10 n$ and
for some $c > 0$ we have
\begin{equation}\label{eq:4b}  
\begin{split}
&\left\{\cap_{i, j\in\{0,1\}}\mathrm{Bad}\,(H^{[k - n + i, k + n - j]},
e^{-cn})\right\} \\
& \hspace{1cm} \cap \left\{ \cap_{i, j\in\{0,1\}}\mathrm{Bad}\,(H^{[l - n +
i, l + n -j]}, e^{-cn})\right\} = \emptyset,
\end{split}%
\end{equation}
where $H^{[m, p]}$ denotes the restriction of $H$ to the interval $[m,p]\cap%
\mathbb{Z}$ with Dirichlet boundary conditions. Then, the corresponding
eigenfunction correlator $Q_I$ of $\eqref{eq:ef-cor}$ for any $I$ decays exponentially,
namely, for any $l \geq 100 n_0$ 
\begin{equation*}  
Q_I(0, l) \leq 16 e^{-\frac{c}{20}l}\equiv 16 e^{-c^{\prime }l}.
\end{equation*}
\end{Lemma}

\begin{Remark}
In the one-dimensional Anderson model, the hypotheses of
Lemma~\ref{lem:bad-res} can be verified using the large-deviation
estimates for the Lyapunov exponent; see, for example,
\cite{JZ} and the references therein. Consequently,
Lemma~\ref{lem:bad-res} combined with Criterion~\ref{cr:EC} yields the Area Law. In this way one
recovers, in the one-dimensional setting, the Area Law result of \cite{EPS}.
\end{Remark}

In addition,
several localisation results and reasonings for one-dimensional finite-difference operators
of second order that are considered below are naturally formulated in terms of 
the Lyapunov exponent. For convenience, we briefly
recall its definition.

Consider the finite-difference  equation
\begin{equation}\label{eq:cau}
u(n+1)+u(n-1)+V_\omega(n)u(n)-\lambda u(n)=0,
\end{equation}
introduce the one-step transfer matrices
\begin{equation}\label{eq:A}
A_\omega(\lambda,n) = 
\begin{pmatrix}
\lambda-V_\omega(n) & -1 \\ 
1 & 0
\end{pmatrix}
, 
\end{equation}
and the corresponding $n$-step transfer matrix 
\begin{equation}\label{eq:Phi}
\Phi_\omega(\lambda,n) = A_\omega(\lambda,n)\cdots A_\omega(\lambda,1), \quad
\begin{pmatrix}
u(n+1)\\
u(n)
\end{pmatrix}
=
\Phi_\omega(\lambda,n)
\begin{pmatrix}
u(1)\\
u(0)
\end{pmatrix}.
\end{equation} 
The Lyapunov exponent is
\begin{equation}\label{eq:le}
\gamma(\lambda)
=
\lim_{n\to\infty}
n^{-1}\log
\|\Phi_\omega(\lambda,n)\|,
\end{equation}
and, by Kingman's subadditive ergodic theorem, the limit exists for almost every $\omega$.

In localisation regimes of one-dimensional operators of second order, 
the Lyapunov exponent determines the natural exponential scale of the spatial 
decay of various spectral characteristics, the Green function in particular. Roughly speaking, one has
\[
|G_\omega(0,n;\lambda)| = O(
e^{-\gamma(\lambda)|n|}), \;|n| \to \infty,
\]
with the precise formulation depending on the model and on the finite- or 
infinite-volume Green function under consideration. 

For many one-dimensional ergodic operators of second order, the positivity of the Lyapunov
exponent is a fundamental ingredient of their localisation theory.
However, for the applications considered in the present survey,
positivity alone is generally insufficient. Depending on the operator in
question,
it must be supplemented by additional quantitative information in
order to verify the hypotheses of Criterion~\ref{cr:EC}. The precise
form of this additional input will be discussed in the next section.

\section{Area Law via various spectral mechanisms}\label{sec:results}

The general Criterion \ref{cr:EC} of the preceding section reduces the 
proof of the Area Law to the verification of sufficiently strong quantitative localisation 
properties of the corresponding one-body Hamiltonian. There is, however, no single spectral 
mechanism that provides the required eigenfunction-correlator estimate for all ergodic operators 
considered here. We therefore discuss separately the several classes  of operators  which require different
mechanisms. 

We begin with operators possessing uniformly localised eigenfunctions (ULE) \eqref{eq:lc}, 
then turn to quasiperiodic operators for which the required estimates follow from 
the property of exponential dynamical localisation, and  finally consider Schr\"odinger operators 
with potentials generated by subshifts of finite type. In this case the spectral input 
of \cite{ADZ1,ADZ} on the  uniform positivity of the Lyapunov exponent and uniform 
large-deviation type estimates was strengthened in \cite{PS} to obtain the exponential 
decay of the eigenfunction correlator. To this
end  the finite-volume Green-function estimates, and a decoupling argument for 
distant symbolic blocks, and Lemma~\ref{lem:bad-res} were developed.

Let us first consider two cases, where the ULE property is the ``best possible''. 
These are  the Maryland  model and the model proposed and analysed in \cite{KPS}, 
both for any dimension $d \ge 1$. Here we have a fairly explicit description 
of the spectrum (see \eqref{eq:evl} and \eqref{eq:efl}), implying a  simple 
proof of the basic bound \eqref{eq:ef-cor-decay} (see Theorem
\ref{th:ule-d-maryland} (iii)). We then consider the one-dimensional operator 
studied in \cite{JKa2}, where a more complex form of the ULE property was obtained.

\subsection{The Maryland model and related operators}\label{subsec:maryland}

\subsubsection{Maryland model}
The model is one of
the few exactly solvable models of the localisation theory of quasiperiodic 
operators which provides the explicit representations of the eigenvalues and eigenfunctions,
as well as explicit formulae for several other spectral and physical quantities, see
\cite[Sections~18.A--18.C]{Pa-Fi:92} and \cite{Si:85}.
 
This explicit structure is of particular
importance for the ULE problem, especially taking into account that in 
contrast with previously considered operators with bounded quasiperiodic 
and limit-periodic potentials, the Maryland model potential, hence
the corresponding operator, is unbounded. Thus, the ULE theorem 
established in \cite{PS} for the Maryland model gives, to the best of 
our knowledge, the first example of the ULE property for an ergodic 
operator on an unbounded spectral interval. In the multidimensional case this interval is the whole real line.

The model is given by the finite-difference operator
\begin{equation}\label{eq:MG}
H_\omega=W+V_\omega
\end{equation}
of   \eqref{eq:lro-a}--\eqref{eq:W} and \eqref{eq:epotg} with the ``Maryland'' potential
\begin{equation}\label{eq:MP}
V_\omega(\mathbf n)
=g\tan\pi\bigl(\omega+\langle \mathbf n,\vec\alpha\rangle\bigr),
\quad
\mathbf n\in\mathbb Z^d,\quad \vec\alpha\in\mathbb T^d,\quad
\omega\in\mathbb T,\quad
g\neq0.
\end{equation}
Thus, the hopping term $W$ is not necessarily the discrete Laplacian. 

We also assume that the frequency vector $\vec\alpha \in\mathbb T^d$ satisfies the 
multidimensional Diophantine condition  (cf.\eqref{eq:Ded} below)
\begin{equation}\label{eq:Dd}
\operatorname{dist}
\bigl(\langle\mathbf n, \vec\alpha\rangle,\mathbb Z\bigr)
\ge
c|\mathbf n|^{-\tau},
\;\;
\mathbf n\in\mathbb Z^d\setminus\{\mathbf 0 \},
\end{equation}
for some  $\mathbf{n}$-independent $c>0$ and $\tau>0$, meaning that  $\vec{\alpha}$ is badly approximated
by vectors with rationally-valued components. 

\smallskip

If $d=1$, then of particular interest is the case where $W$ is the discrete Laplacian, 
hence, $H_\omega$ is the discrete Schr\"odinger operator \eqref{eq:Sch} with potential \eqref{eq:MP} for $d=1$
\begin{equation}\label{eq:MG1}
(H_\omega\psi)(n)
=\psi(n+1)+\psi(n-1)
+
g\tan\pi(\omega+n\alpha)\psi(n),
\;
\end{equation}
and $\alpha$ satisfies a
more delicate arithmetic than \eqref{eq:Dd} condition $ 0 \leq \beta(\alpha) \leq \beta_0$ where
\begin{equation}\label{eq:beta}
\beta(\alpha)
:=\limsup_{n\to\infty}
\; n^{-1}(-\log \operatorname{dist}(n\alpha,\mathbb Z)),
\end{equation}
and $\beta_0=\beta_0(g)>0$ is sufficiently small. This allows both Diophantine and weak Liouville frequencies.

\medskip
Under these conditions the spectral structure of the Maryland 
model \eqref{eq:MG} -- \eqref{eq:beta} is as follows, see \cite{Ji-Li:17,Pa-Fi:92,Si:85}.
\begin{itemize}
\item[(i)]
$d \ge 1$, see \eqref{eq:MG} -- \eqref{eq:Dd} and \eqref{eq:W}. Then the spectrum is
\begin{equation}\label{eq:spR}
\sigma(H_\omega)=\mathbb R,
\end{equation}
and the spectral type is pure point and simple. 
The eigenvalues and normalised eigenfunctions can be labelled by
$\mathbf l\in\mathbb Z^d$ so that there exist    $\lambda_0:\mathbb T\to\mathbb R$ 
and $\psi_{\mathbf 0}:\mathbb Z^d\times\mathbb R\to\mathbb C$
independent of $\omega$ and $\mathbf l$, such that the eigenvalues and
eigenfunctions  of $H$
are
\begin{equation}\label{eq:evl}
\{\lambda_\mathbf{l}\}_{\mathbf{l} \in \mathbb{Z}^d}, \; 
\lambda_\mathbf{l}=\lambda_0(\omega+\langle \mathbf l, \vec\alpha\rangle),
\end{equation}
and the eigenfunctions are
\begin{equation}\label{eq:efl}
\{\psi_\mathbf{l}\}_{\mathbf{l} \in \mathbb{Z}^d}, \;\psi_\mathbf{l}
=\{\psi_\mathbf{l}(\mathbf{n})\}_{\mathbf{n} \in \mathbb{Z}^d}, \;
\psi_\mathbf{l}(\mathbf{n})=\psi_{\mathbf 0}(\mathbf{n}-\mathbf{l},\lambda_\mathbf{l}),
\end{equation}

\medskip

\item[(ii)]
$d=1$, see \eqref{eq:MG1} -- \eqref{eq:beta}. Let 
$\gamma(\lambda,g)$ be the Lyapunov exponent \eqref{eq:le} of the 
corresponding finite-difference equation of second order \eqref{eq:cau}, and 
let $\varepsilon_0\ge0$ be determined by
(see \cite{Pa-Fi:92}, Section 18.C)
\[
\begin{cases}
\gamma(\varepsilon_0,g)=\beta(\alpha), \qquad
& \text{if }\beta(\alpha)\ge\gamma(0,g),\\[3pt]
\varepsilon_0=0, \qquad
& \text{if }\beta(\alpha)\le\gamma(0,g).
\end{cases}
\]

If $\varepsilon_0=0$, the pure-point spectrum of the operator \eqref{eq:MG1} occupies the whole spectral axis,
cf. \eqref{eq:spR}, and eigenvalues and
eigenfunctions  admit the description which is the $d=1$ version of \eqref{eq:evl} -- \eqref{eq:efl}. 

If $\varepsilon_0>0$, then the spectrum is again  the whole spectral axis, but it is
pure point and simple only on the two unbounded
intervals
\begin{equation}\label{eq:spp}
\sigma_{pp}(H)=(-\infty,-\varepsilon_0)\cup(\varepsilon_0,\infty),
\end{equation}
while on the central interval
$[-\varepsilon_0,\varepsilon_0]$ the spectrum is purely singular continuous
\[
\sigma_{sc}(H)=[-\varepsilon_0,\varepsilon_0].
\]
The eigenvalues and
eigenfunctions  of \eqref{eq:MG1} admit on \eqref{eq:spp} the description 
which is the $d=1$ version of \eqref{eq:evl} -- \eqref{eq:efl}.
\end{itemize}

\smallskip 
 
The proof of the above spectral structure is based  on a simple observation
that  convolution operators and multiplication operators by $e^{2\pi i <\vec{\alpha}, \mathbf{n}>}$ 
with varying $\vec{\alpha}$ form two commutative algebras. This simple and generally 
not too useful observation proves to be quite efficient in this case of the Maryland model  
(see \cite[Sections~18.A--18.C]{Pa-Fi:92}).

\medskip

These spectral considerations identify the regions in which the ULE
mechanism can lead to the Area Law. We now state the uniform localisation
result for the Maryland model proved in \cite{PS}. The result is new in both the one- 
and multidimensional cases and
provides the first examples of ULE for unbounded ergodic operators. 

\begin{Theorem}
\label{th:ule-d-maryland}
Let $H_\omega$ be the Maryland model described by 
\eqref{eq:MG} -- \eqref{eq:beta}.
\begin{itemize}
\item[(i)]
$d\geq 1$, see \eqref{eq:MG} -- \eqref{eq:Dd} and \eqref{eq:W}. Then $H_{\omega}$ 
has uniformly localised eigenfunctions $\{\psi_{\mathbf{l}}(\omega)\}_{\mathbf{l}
\in \mathbb{Z}^d}$ on the whole spectrum 
\[
\sigma(H_\omega)=\sigma_{pp}(H_\omega)=\mathbb R,
\]
so that we have for $\psi_{\mathbf{l}}(\omega)=\{\psi_\mathbf{ l}(\omega,\mathbf n)\}_
{\mathbf{n}
\in \mathbb{Z}^d}$
\begin{equation}\label{eq:ulema}
|\psi_{\mathbf l}(\omega,\mathbf n)|
\le
Ce^{-c|\mathbf n-\mathbf {l}|},
\;
\mathbf n\in\mathbb Z^d,
\end{equation}
with $C<\infty$ and $c>0$ independent of $\omega \in \mathbb{T}$ and $\mathbf{l},\mathbf{n}
\in\mathbb Z^d $ (cf. \eqref{eq:lc}), i.e., with the set $\{\mathbf{n}_{\mathbf{l}}\}$
of \eqref{eq:lc}
coinciding with  $\mathbb{Z}^d$;

\smallskip

\item[(ii)]
$d=1$,  see \eqref{eq:MG1} -- \eqref{eq:beta}. Then $H_{\omega}$ has uniformly 
localised eigenfunctions   $\{\psi_{l}(\omega)\}_{l
\in \mathbb{Z}} $ on\[
\sigma_{pp}(H_\omega)=(-\infty,-\varepsilon_0)\cup(\varepsilon_0,\infty),
\]
so that
we have here  the $d=1$ version of \eqref{eq:ulema}.

\smallskip
\item[(iii)] The Area Law is valid for Fermi energy $\varepsilon_F$ 
belonging to the whole $\mathbb{R}$ in the multidimensional case (i) and for
$\varepsilon_F \in (-\infty, -\varepsilon_0)$ in the one-dimensional case
(ii). 
\end{itemize}
\end{Theorem}

\medskip \noindent

The proof of the theorem relies on a representation of the eigenfunctions 
$\psi _{\mathbf{l}}:\mathbb{Z}^{d}\rightarrow \mathbb{C}$ of \eqref{eq:efl} as Fourier integrals of 
functions $\widehat{\psi} _{\mathbf{l}}:\mathbb{T}^{d}\rightarrow \mathbb{C}$ 
obtained via a rather complex procedure (see \cite{Pa-Fi:92}, Section~18.B--18.C). 
It is shown that this procedure admits an analytic continuation to the complex neighbourhood
$$
\mathbb{T}_{\rho }^{d}
=\{z=\{z_{1},\ldots ,z_{d}\}\in \mathbb{C}^{d}:e^{-\rho }\leq |z_{t}|\leq e^{\rho },\;\rho >0,\;t=1,\ldots ,d\}
$$ 
of $\mathbb{T}^d$ for some $\rho >0$.  

This makes it possible to deform the ``contour'' $\mathbb{T}^{d}$ into a contour lying 
within $\mathbb{T}_{\rho }^{d}$, hence, leading to the exponential estimate \eqref{eq:ulema}.

\begin{remark}
The proof of validity of an analogue of Assertion (iii) of the theorem in more general
 than \eqref{eq:ulema} cases of the ULE property \eqref{eq:lc} requires an
additional argument  (see Section~\ref{sec:tools} for the corresponding discussion). 
The argument can
be extracted from the works \cite{JK} and \cite{Ha:16}. The first provides
a method to deal with general distributions of localisation centres in \eqref{eq:lc}.
According to the second the parameters $C$ and $c$ in \eqref{eq:lc} are independent
of $\omega$ if the corresponding dynamical system \eqref{eq:psp} -- \eqref{eq:auto}
is minimal. Quasiperiodic operators generated by irrational
rotations of the torus provide the typical examples of this situation.
\end{remark}

\subsubsection{Related operators}
The Maryland model manifests particularly clearly how an explicit
description of the pure point spectrum  can lead to uniform control of the
eigenfunctions. Related ideas have subsequently been used for  other
quasiperiodic and limit-periodic operators for which the eigenfunctions admit sufficiently
precise representations or parametrisations. We will present two such
classes of operators whose  quasiperiodic potentials have a sampling function 
that is monotone on one period (see \eqref{eq:Hod} and \eqref{eq:Ho1} below), 
 so that a large part of resonance phenomena is suppressed. 

Although the corresponding
localisation theorems for these operators are obtained by different methods than those for the
Maryland model, the resulting
uniform eigenfunction estimates can again be converted into exponential
decay of the eigenfunction correlator through the localisation-centre
argument of Theorem~\ref{th:ule-d-maryland}(iii) or \cite{JK}.

We first consider 
a multidimensional operator considered by Kachkovskiy,
Parnovski and Shterenberg \cite{KPS}. This is
the multidimensional discrete Schr\"odinger operator
\eqref{eq:Sch} with potential (cf. \eqref{eq:epotg} and \eqref{eq:MP})
\begin{equation}\label{eq:Vgv}
V(\omega,\mathbf n)
=
gv(\omega+\langle\mathbf n, \vec\alpha\rangle),
\;
\mathbf n\in\mathbb Z^d, \quad \vec\alpha\in\mathbb T^d,
\end{equation}
where $g>0$,  
$v:\mathbb R\to\mathbb R$ is $1$-periodic,  satisfies the
$\xi$-H\"older monotonicity condition
\begin{equation}\label{eq:Hod}
v(y)-v(x)
\ge
(y-x)^\xi,
\qquad
\xi\ge1,
\quad
0\le x\le y<1,
\end{equation}
and the frequency vector $\vec\alpha$ satisfies the weak Diophantine condition
(cf. \eqref{eq:Dd} and \eqref{eq:beta})
\begin{equation}\label{eq:Ded}
\operatorname{dist}
(\langle\mathbf n, \vec\alpha\rangle,\mathbb Z)
\geq
\exp\!\left(
-\rho|\mathbf n|^{1/(1+\mu)}
\right),
\quad
\mathbf n\in\Z^d\setminus\{\mathbf 0\} \qquad
\rho,\mu>0.
\end{equation}
By using an appropriately organised version of the large coupling perturbation theory,
it was shown in  \cite{KPS} that if $g$ is large enough, then the same description \eqref{eq:evl} -- \eqref{eq:efl} 
of the spectrum and eigenfunctions as that of the Maryland model holds in this case also. 
Then, Assertion (iii) of Theorem \ref{th:ule-d-maryland} applies and yields the validity of the Area Law for
 $\varepsilon_{F} \in \mathbb{R}$.

\smallskip

We next consider the one-dimensional case 
\eqref{eq:Sch} with the $d=1$ version of potential \eqref{eq:Vgv} 
where $g>0$ is sufficiently large,
$v:\mathbb R\to\mathbb R, \; v(0)=0,
\;
v(1-)=1$ is $1$-periodic, continuous on $[0,1)$,
 Lipschitz monotone (cf. \eqref{eq:Hod})
\begin{equation}\label{eq:Ho1}
a_-(y-x)
\le
v(y)-v(x)
\le
a_+(y-x),
\;
0\le x\le y<1, \ a_+, a_- > 0,
\end{equation}
and $\alpha$ satisfies the Diophantine condition
\eqref{eq:Dd} for $d=1$.

For this class of operators, Jitomirskaya and Kachkovskiy
\cite{JKa2} established the ULE property \eqref{eq:lc} which,
combined with the  localisation centre argument by
Jitomirskaya and Kr\"uger \cite{JK}, implies the conclusions of
Criterion~\ref{cr:EC}, hence, the validity of the Area Law for the above one-body
Hamiltonian. 

One more class of ergodic operators consists of multidimensional 
Schr\"odinger operators \eqref{eq:Sch} 
with the limit-periodic case of potential \eqref{eq:epotg},
where $\Omega$ of \eqref{eq:psp} is a Cantor group admitting a minimal $\mathbb{Z}^d$-action
by translations, 
$T:\mathbb{Z}^d\times\Omega\to\Omega$,
and 
$v\in C(\Omega,\mathbb{R})$.

Damanik and Gan \cite{Da-Ga:11,Da-Ga:13} proved the ULE property \eqref{eq:lc} with
$\omega$-independent constants $C$ and $c$ for a class of such limit-periodic operators. 
This and the results of \cite{JK} lead
to the bound \eqref{eq:ef-cor-decay}, hence to the Area Law via 
Criterion~\ref{cr:EC}.

\subsection{Quasiperiodic operators beyond ULE}\label{subsec:qp}

We consider now the quasiperiodic operators for which the hypotheses of
Criterion~\ref{cr:EC} follow from localisation results that do not proceed
through the ULE property \eqref{eq:lc}. The presently available examples of this type are the one-dimensional supercritical almost Mathieu operator
and a multidimensional long-range hopping analogue of the almost Mathieu operator. 

The first operator is 
\begin{equation}\label{eq:AMO}
(H_\omega\psi)(n)
=
\psi(n+1)+\psi(n-1)
+
2g\cos 2\pi(\omega+n\alpha)\psi(n),
\; n\in\mathbb Z,
\end{equation}
where
$
\omega \in\mathbb T,
\;\;
|g|>1$. Assuming that   $\alpha \in\mathbb T $ 
satisfies 
\[
0\leq\beta(\alpha)<\beta
\]
with 
$\beta(\alpha)$ of \eqref{eq:beta} and a  suitable $\beta>0$, Jitomirskaya and Kr\"uger \cite{JK} proved
the $d=1$ version of exponential dynamical localisation in 
expectation \eqref{eq:exp-dyn-loc}. 

\smallskip

The second operator is the finite-difference operator
$H_\omega=W+V_\omega$
of \eqref{eq:lro-a}--\eqref{eq:W}, with
potential (cf. \eqref{eq:MG} -- \eqref{eq:MP})
\[
V_\omega(\mathbf n)
=
g\cos 2\pi\bigl(\omega+\langle\mathbf n, \vec\alpha\rangle\bigr),
\;\;
\mathbf n\in\mathbb Z^d,\;\; \omega\in\mathbb T.
\]
Thus the hopping term $W$ is a discrete convolution, subject to the
exponential decay assumption \eqref{eq:W}. 

The frequency vector $\vec\alpha$ satisfies
the following version of the Diophantine condition (cf. \eqref{eq:Dd})
\begin{equation}
DC_d
=
\bigcup_{\kappa>0,\;\tau>d-1}
\left\{
\vec\alpha\in\mathbb T^d:
\operatorname{dist}
\bigl(\langle\mathbf n, \vec\alpha\rangle,\mathbb Z\bigr)
>
\kappa |\mathbf n|^{-\tau},
\;
\mathbf n\in\mathbb Z^d\setminus\{0\}
\right\}.
\label{eq:DC_d}
\end{equation}
For every $\vec{\alpha}\in DC_d$, Ge, You and Zhou \cite[Theorem~1.2]{GYZ}
proved also that this ergodic operator exhibits exponential
dynamical localisation in expectation \eqref{eq:exp-dyn-loc}  if $g$ is sufficiently large. 

\smallskip

 For the present purposes, both above results contain slightly more information
than their stated dynamical-localisation conclusions \eqref{eq:exp-dyn-loc}. 
Namely, in the proofs of
\cite{GYZ} and \cite{JK}, the dependence on the time evolution enters only
through the pointwise bound
$
|e^{-itx}|\leq1.
$
Thus, the same arguments apply with $e^{-itx}$ replaced by an arbitrary
bounded continuous function $\phi$ satisfying $\|\phi\|_\infty\leq1$ of \eqref{eq:ef-cor}.
Taking the supremum over such functions gives precisely the exponential
decay estimate for the  correlator
\eqref{eq:ef-cor-decay}. Hence Criterion~\ref{cr:EC} applies to both
classes of ergodic operators, yielding the Area Law throughout the corresponding localisation regimes.

\subsection{Schr\"odinger operators generated by subshifts of finite type}\label{subsec:subshift}

We now turn to one-dimensional discrete Schr\"odinger operators \eqref{eq:Sch} whose potentials 
are generated by hyperbolic dynamical systems. 
Two famous examples of such systems are the doubling map and Arnold's cat map,
\begin{equation}\label{eq:dmap}
T_2\omega=2\omega\pmod 1,\qquad
T_{\mathrm{cat}}(\omega_1,\omega_2)
=(2\omega_1+\omega_2,\omega_1+\omega_2)\pmod 1.
\end{equation}
Both admit finite Markov partitions and symbolic codings by subshifts of finite type \cite{Adler,Br-St:02}. 
For the doubling map, the two-sided coding refers to its natural extension. Unlike quasiperiodic potentials, 
whose long-range correlations reflect the arithmetic of irrational rotations, 
these potentials have statistical properties closer to those of random potentials. 
They thus provide a natural intermediate class between quasiperiodic and random operators.

Let $\mathcal A=\{1,\ldots,k\}$, $k\ge 2$, be a finite alphabet and let $A$ be a $k\times k$ 
matrix with entries in $\{0,1\}$. The corresponding dynamical 
system \eqref{eq:psp} --  \eqref{eq:auto} 
known as associated subshift of finite type consists of the following event space and its shift transformation
\begin{equation}\label{eq:sft-space}
\Omega_A=\{\omega\in\mathcal A^{\mathbb Z}:
A_{\omega_n\omega_{n+1}}=1\text{ for all }n\in\mathbb Z\},
\qquad (T\omega)_n=\omega_{n+1}.
\end{equation}
Its  $\sigma$-algebra is generated by the cylinders
\begin{equation}\label{eq:sft-cylinders}
[s;a_0,\ldots,a_m]
=\{\omega\in\Omega_A:\omega_{s+j}=a_j,\quad 0\le j\le m\}.
\end{equation}
It is metrised by
\begin{equation}\label{eq:sft-metric}
d(\omega,\widetilde\omega)=e^{-N(\omega,\widetilde\omega)},\qquad
N(\omega,\widetilde\omega)=\min\{|j|:\omega_j\ne\widetilde\omega_j\},
\end{equation}
for $\omega\ne\widetilde\omega$, and $d(\omega,\omega)=0$.

{
To define the probability measure on $\Omega_A$, 
introduce the stochastic matrix  $P=\{P_{ab}\}_{a,b \in
\mathcal{A}}$ with $P_{ab}>0$ precisely when $A_{ab}=1$ and consider a    
finite-state
Markov chain having $P$ as its transition matrix. If $\mathbf p=\{p_a\}_{\{a
\in \mathcal{A}}$ is its stationary distribution, i.e., $ \mathbf pP=\mathbf p$, then the corresponding shift
invariant Markov measure is determined by}
\begin{equation}\label{eq:markov-measure}
\mathbf P_P([0;a_0,\ldots,a_m])
=p_{a_0}P_{a_0a_1}\cdots P_{a_{m-1}a_m},\qquad \mathbf pP=\mathbf p.
\end{equation}
For irreducible $A$ of \eqref{eq:sft-space}, hence, $P$, this  Markov measure is ergodic. If $T$ has a fixed point, 
$P$ is also aperiodic, and the corresponding chain is exponentially mixing \cite[Theorem~4.9]{LPW}. 
In terms of events $E,F$ determined by coordinate blocks separated by a distance $r$, this gives
\begin{equation}\label{eq:sft-mixing}
\left|\mathbf P_P(E\cap F)-\mathbf P_P(E)\mathbf P_P(F)\right|
\le C e^{-cr},\qquad C<\infty,\quad c>0.
\end{equation}

Given a real-valued sampling function $v:\Omega_A\to\mathbb R$, we introduce
the ergodic potential as (cf. \eqref{eq:epotg})
\begin{equation}\label{eq:sft-potential}
V_\omega(n)=v(T^n\omega),\qquad n\in\mathbb Z.
\end{equation}
A function $v$ is locally constant if it depends on a fixed finite block of coordinates. It is 
$\alpha$-H\"older, $0<\alpha\le1$, if
\begin{equation}\label{eq:sft-holder}
|v(\omega)-v(\widetilde\omega)|
\le C_vd(\omega,\widetilde\omega)^\alpha.
\end{equation}
Thus, the values of $v$ on $[-N,N]$ determines it up to an error $O(e^{-\alpha N})$; 
for locally constant $v$, the determination is exact for sufficiently large $N$.

We impose the following

\medskip
\textbf{Assumptions}:
\begin{itemize}
\item[(a)] the adjacency matrix $A$ in \eqref{eq:sft-space} is irreducible;
\item[(b)] $\mathbf P$ is a $T$-ergodic probability measure with support $\Omega_A$ of \eqref{eq:sft-space};
\item[(c)] $\mathbf P$ has the bounded distortion property: there exists $C_{\mathrm{bd}}\ge1$ such that
\begin{equation}\label{eq:bounded-distortion}
C_{\mathrm{bd}}^{-1}\le
\frac{\mathbf P(E\cap F)}{\mathbf P(E)\mathbf P(F)}
\le C_{\mathrm{bd}}
\end{equation}
whenever $E,F$ are cylinders supported on disjoint coordinate intervals and have nonempty intersection;
\item[(d)] $T$ of \eqref{eq:sft-space} has a fixed point;
\item[(e)] $v$ of \eqref{eq:sft-potential} is not $\mathbf P$-almost surely constant and either locally constant or satisfies 
\eqref{eq:sft-holder} with $\|v\|_\infty=\sup_{\omega\in\Omega_A}|v(\omega)|$ sufficiently small.
\end{itemize}
The smallness requirement in (e) ensures the so-called global fibre-bunching condition used in \cite{ADZ}.

For the Schr\"odinger operator \eqref{eq:Sch} with potential
\eqref{eq:sft-potential}, it is proved in \cite{ADZ1,ADZ}
that there exist a finite exceptional set $\mathcal F_v$ and a finite
interval $J$ containing its spectrum and such that, for every $\eta>0$,
the Lyapunov exponent \eqref{eq:le} is uniformly positive
and the transfer matrices \eqref{eq:Phi} satisfy a uniform
large-deviation estimate on
\begin{equation}\label{eq:sft-energy-set}
J_\eta=J\setminus B(\mathcal F_v,\eta),\qquad
B(\mathcal F_v,\eta)
=\{\lambda:\operatorname{dist}(\lambda,\mathcal F_v)<\eta\}.
\end{equation}

We have then \cite{PS}

\begin{Proposition}\label{th:subshift-ec-survey}
Assume Assumptions (a)--(e) above.
Then for every compact  set $I\subset J\setminus\mathcal F_v$,
the eigenfunction correlator \eqref{eq:ef-cor} satisfies
the exponential estimate \eqref{eq:ef-cor-decay}.
Consequently, $H_\omega$ exhibits exponential dynamical localisation
in expectation \eqref{eq:exp-dyn-loc} on $I$.
\end{Proposition}

\begin{Theorem}\label{cor:subshift-area-survey}
Assume Assumptions (a)--(e) above.
Then there exists $\varepsilon_0>\varepsilon_-$ such that,
for $\varepsilon_F\in(\varepsilon_-,\varepsilon_0]$,
the expectation of the entanglement entropy \eqref{eq:ee}
obeys the Area Law \eqref{eq:area-law-criterion}.
\end{Theorem}

Note that for every interval $I\subset J_\eta$, the localisation results
of \cite[Remark~2.13]{ADZ}  include the almost-sure bound
\begin{equation}\label{eq:sft-adz-dl}
\sup_{t\in\mathbb R}
\left|
\left\langle
\delta_n,e^{-itH_\omega}\chi_I(H_\omega)\delta_m
\right\rangle
\right|
\le C_{\omega,b,\epsilon}\,e^{\epsilon|m|}e^{-b|n-m|},
\qquad m,n\in\mathbb Z,
\end{equation}
were $\epsilon>0$ and
$0<b<\inf_{\lambda\in I}\gamma(\lambda)$ are arbitrary,
and the coefficient $C_{\omega,b,\epsilon}$ depends on the realisation.
This and the exponentially growing factor $e^{\epsilon|m|}$ makes this estimate unappropriate  for  controlling the spatial sums entering the entropy bound
uniformly in the block size.
Thus, the bound \eqref{eq:sft-adz-dl} does not implies exponential dynamical localisation
in expectation \eqref{eq:exp-dyn-loc}, or the exponential estimate
\eqref{eq:ef-cor-decay} for the eigenfunction correlator
\eqref{eq:ef-cor} needed to obtain the Area Law
\eqref{eq:area-law-criterion} through Criterion~\ref{cr:EC}.

The proof of above proposition and the theorem  in \cite{PS} uses the general criterion of
Lemma~\ref{lem:bad-res}, which gives exponential decay of the
eigenfunction correlator for bounded ergodic one-dimensional
Schr\"odinger operators under a condition on four variants of
finite-volume restrictions, see \eqref{eq:4b}.
For subshifts of finite type, the idea is to approximate the sampling function
 $v$ of \eqref{eq:sft-potential} by
locally constant functions, so that distant restrictions depend
on separated symbolic blocks and can be treated as approximately
independent.
For ergodic Markov measures \eqref{eq:markov-measure},
this approximate independence is quantified by exponential mixing
\eqref{eq:sft-mixing}; the bounded distortion assumption (c)
provides the conditional probability control needed in the general case.
To carry out this idea, however, one needs a large-deviation estimate for
the Lyapunov exponent  controlling all subintervals of each box simultaneously.

The required strengthening of the uniform large deviations estimates  of \cite{ADZ} is proved
in \cite{PS}:
\begin{equation}\label{eq:sft-uld-subintervals}
\mathbf P\!\left\{
\begin{array}{c}
\exists\,k,l,\quad -n\le k\le l\le n:\\[2pt]
\left|\log\|\Phi_{T^k\omega}(\lambda,l-k)\|
-(l-k)\gamma(\lambda)\right|\ge\epsilon n
\end{array}
\right\}
\le C_{\epsilon,v,I}e^{-c_{\epsilon,v,I}n},
\qquad \lambda\in I.
\end{equation}
The coefficients  are uniform in $\lambda\in I$, and the error
is measured on the scale of the whole box.

The Green function  estimates translate this bound into the control of the
bad-energy sets \eqref{eq:bad}, accounting for the approximation
error and resonant energies.
The resulting sets are unions of $O(n)$ intervals, and an
intersection in Lemma~\ref{lem:bad-res} 
must contain an endpoint of one of these intervals.
Fixing the potential near one box therefore leaves only $O(n)$
prescribed energies to check in a distant box.
Approximate independence allows for the fixed-$\lambda$ probability bound
to be applied to these checks.
Their number grows only linearly, whereas the probability bound
decays exponentially, so the resulting bound remains exponentially small.
Combined with the argument of Lemma~\ref{lem:bad-res},
valid for any interval, this gives \eqref{eq:ef-cor-decay}
for the eigenfunction correlator \eqref{eq:ef-cor}.

The complete proof is given in \cite{PS}.
The bound \eqref{eq:ef-cor-decay} therefore yields both exponential
dynamical localisation in expectation \eqref{eq:exp-dyn-loc}
and, through Criterion~\ref{cr:EC} applied on a bottom spectral
interval, the Area Law \eqref{eq:area-law-criterion}.

\section{Conclusion}
We reviewed above the conditions and proofs of validity of the  Area Law, the large-size asymptotic form of the entanglement entropy of free lattice fermions 
in terms of spectral properties of the one-body Hamiltonian $H$ of fermions. 
Specifically, we assumed that $H$ is the discrete Schr\"odinger operator with potential 
generated by deterministic ergodic systems (irrational windings of the torus and subshifts 
of finite type). It was explained that in view of existing results the problem reduces 
to the proofs of exponentially decaying bounds on the entries of the spectral 
(Fermi) projection \eqref{eq:exp-decay} or the eigenfunction correlator \eqref{eq:ef-cor-decay} 
of the operator in question. The bounds are ultimately related to the Anderson localisation, a strong quantitative form of pure point spectrum of the operator. 

In other words, a common feature of the presently known rigorous proofs 
of the Area Law for the free fermions in ergodic field 
is as 
follows. The entropy estimate itself is largely insensitive to the mechanism producing 
the ``strong localisation'' and the exponential bounds \eqref{eq:exp-decay} or  \eqref{eq:ef-cor-decay}. 
What varies from one class of operators to another is the spectral argument needed 
to verify the relevant decay hypothesis. 

For random operators, general powerful  methods such 
as multiscale analysis and the fractional-moment method are available. For deterministic ergodic 
operators, no similar universal methods exist: quasiperiodic (bounded and unbounded), 
limit-periodic, and hyperbolically generated potentials require substantially different spectral inputs. 

We considered several classes of operators. Two of them are of particular 
interest. The first is the Maryland model in any dimension, where established
in \cite{PS} uniform 
localisation of eigenfunctions on an infinite spectral interval is the corresponding mechanism, providing a direct
route to the decay of the Fermi projection and the eigenfunction correlator. 

The second consists of the discrete Schr\"odinger operators with potentials generated by
subshifts of finite type. Here also the existing results on spectral localisation needed 
to be considerably strengthened to obtain exponential decay of the eigenfunction correlator and, 
consequently, exponential dynamical localisation in expectation
and the Area Law.

We believe that the present survey brings some order to this small but active 
body of rigorous results, thereby helping to convert it into a unified mathematical picture. 
We review the proofs of the Area Law for free fermions with ergodic one-body Hamiltonians, 
explain the role of the Fermi projection and the eigenfunction correlator, and describe 
how the required decay estimates arise in interesting classes of ergodic deterministic 
Hamiltonians. Particular attention is paid to the point at which existing localisation results 
already suffice and to the point at which new spectral arguments become necessary. The 
precise Area-Law statements were presented only after the corresponding general spectral 
mechanism had been made clear, so that the individual models appear not as an 
unrelated list of examples but as distinct realisations of the same spectral criterion.

The theory remains far from complete.
All currently available proofs in this setting rely on strong quantitative
localisation, but no converse principle is known.
In particular, it remains open whether the Area Law \eqref{eq:al}
implies ``strong'' localisation of the corresponding one-body Hamiltonian
in the relevant spectral region.
Another question is whether, in the absence of transparency energies as in \cite{Mu-Co:20}, 
the sufficient localisation assumptions
can be weakened to pure point spectrum with exponentially decaying
eigenfunctions.
It is also unclear which finer spectral characteristics govern
the transition between the Area Law \eqref{eq:al}
and the Enhanced Area Law \eqref{eq:eal}.
These questions suggest that the relation between entanglement entropy
and spectral theory is considerably subtler than the presently known
sufficient criteria indicate.

\end{document}